# Effect of AC current annealing on the microstructure, magnetism and magnetoimpedance of CoFeSiBNb$_3$ microfibers

Jingshun Liu [a,*], Feng Wang [a], Meifang Huang [b], Yun Zhang [c], Congliang Wang [a], Ze Li [a], Hongxian Shen [d], and Manh-Huong Phan [e]

[a]School of Materials Science and Engineering, Inner Mongolia University of Technology, Hohhot 010051, People's Republic of China

[b]Center for Green Innovation, School of Mathematics and Physics, University of Science and Technology Beijing, Beijing 100083, People's Republic of China

[c]Instrumentation Engineering Faculty, Belarusian National Technical University, Minsk 220013, Belarus

[d]School of Materials Science and Engineering, Harbin Institute of Technology, Harbin 150001, People's Republic of China

[e]Department of Physics, University of South Florida, Florida 33620, USA

**Abstract**

This paper systematically studies the changes in the microstructure and magnetic properties of CoFeSiBNb$_3$ metallic microfibers before and after AC annealing. The influence of current intensity on the magneto-impedance (MI) effect of the microfibers was analyzed and the microstructure changes of the microfibers before and after annealing were explored by means of high-resolution transmission electron microscopy. Using this data, the mechanism of current annealing to improve the MI characteristics was further revealed. The results show that the surface of the CoFeSiBNb$_3$ metallic microfibers after AC current annealing is smooth and continuous; its general magnetic properties and MI ratio increase at first with current intensity and then decrease at higher intensities. Specifically, the 140-mA AC annealed metallic microfiber shows excellent magnetic properties with $M_s$, $\mu_m$, $H_c$, and $M_r$ values reaching 94.38 emu/g, 0.4326, 34.36 Oe, and 13.94 emu/g, respectively. With an excitation frequency of $f$ = 1

---

[*]Corresponding author.
E-mail address: jingshun_liu@163.com (Jingshun Liu)

MHz, the wire's $[\Delta Z/Z_{max}]_{max}$, $\xi_{max}$, $H_k$, and $H_p$ reached 216.81%, 31.04 %/Oe, 1 Oe, and 2 Oe, respectively. During the current annealing process, Joule heat eliminates residual stresses in the microfibers while forming atomically ordered micro-domains, which improves the degree of its organizational order. Meanwhile, a stable toroidal magnetic field is generated, which promotes the distribution of the magnetic domain structure of the microfibers, thereby improving the MI effect.

## 1. Introduction

With the development of information technology, the demand for information and signal conversion rapidly increases. As a virtual information conversion device, magneticسensors have been widely employed in biomedical, geomagnetic detection, aerospace, and military industries [1-4]. Compared with traditional sensors, magneto-impedance (MI) based magnetic sensors have advantages of high sensitivity and miniaturization [5,6]. Moreover, among sensitive magnetic materials, quasi-one-dimensional amorphous metallic microfibers are the first choice as sensing materials for MI sensors due to their excellent intrinsicسensitivity, low hysteresis loss, and high magnetic permeability [7,8]. Meanwhile, the skin effect exhibited by Co-based metallic microfibers is beneficial in improving their MI performance [9]. Along with its geometric symmetry and short-range ordered amorphous structure, Co-based microfibers are suitable microscale materials to enhance sensor integration and achieve device miniaturization [10-12].

The giant magneto-impedance (GMI) effect refers to the phenomenon of a significant change in internal impedance of a magnetic material with a small applied external magnetic field under the excitation of a high-frequency alternating current (AC) [13,14]. The most popular amorphous metallic microfibers for this purpose are Co-based [15], Fe-based [16], or Co-Fe-based [17]. In general, Co-based metallic microfibers are usually more likely to develop a "core-shell" structure than others due to the combined action of stress and negative hysteresis during formation, resulting in an alignment of magnetic moments in the circumferential direction [18-20]. Additionally, after annealing, Co-based metallic microfibers exhibit a higher GMI change rate and greater sensitivity to magnetic field than Fe-based metallic microfibers [21]. Typically, the internal magnetic domain structure of the microfibers can be adjusted by changing its chemical composition and obtaining metallic microfibers with excellent magneto-impedance performance [22]. For example, Nb, a good superconductor, has a high magnetic penetration depth that effectively optimizes the magnetic domain distribution. Shen et al. [23] investigated the effects of Nb doping on the GMI properties of Co-based metallic microfibers and found that the GMI properties tended to increase and then decrease as B was replaced by Nb. Introducing Nb to the microfiber composition disturbs its initial equilibrium structure. When Nb is supplied in excess, the GMI effect of the microfiber is reduced and more residual stress may be introduced. According to Thiabgoh et al. [24], multi-step Joule annealing can enhance the high-frequency characteristics of Co-based metallic microfibers with progressive relaxation of their microclusters and magnetic domain structure. However, this annealing method is laborious and only suitable for small magnetic fields and microsensors. Furthermore, preparing metallic microfibers requires rapid cooling, resulting in a large amount of residual stress remaining inside the microfibers and insufficient GMI characteristics [25-27].

Meanwhile, Zhou et al. [28] demonstrated that the MI effect of Co-based amorphous thin ribbons could be affected by both annealing procedure and cooling conditions. It was found that the Joule heat produced by the current modulation treatment not only modified the toroidal domain structure but also removed residual stresses, thereby significantly enhancing the GMI effect. Therefore, proper annealing treatment of metallic microfibers can effectively reduce the influence of residual stress. Similarly, Tiberto et al. [29] carried out Joule annealing and ordinary annealing on microfibers and found that Joule heating improved the MI ratio. In contrast, the effect of traditional annealing technology was not obvious and, more importantly, any internal crystallization of the amorphous microfibers would worsen the MI effect. Therefore, if the current annealing process is not properly designed, large areas of crystallization inside the microfiber can occur, increasing the magnetic crystal anisotropy. This increases the driving force required for the rotation of the magnetic moment, which seriously affects the alignment of the magnetic moment and reduces the GMI effect. As a result, in order to establish a good match between the amorphous microfiber structure and the magnetic impedance, current parameters must be strictly regulated throughout the current annealing process for magnetically sensitive materials.

This study aims to compare and investigate the influence of the microstructure and magnetic properties of CoFeSiBNb$_3$ microfibers before and after AC annealing treatment. We systematically analyze the effect of current intensity on the MI effect of the microfibers to reveal the mechanism of the impact of nano-clusters and ordered micro-regions on the MI behavior of CoFeSiBNb3 with the help of high-resolution transmission electron microscopy, further providing a theoretical basis for the preparation of sensitive materials for high-performance MI sensors.

## 2. Experimental details

The nominal composition of the metallic microfibers used in this study is $Co_{69.25}Fe_{4.25}Si_{13}B_{10.5}Nb_3$ (at.%). The raw material is obtained by arc melting to obtain the master alloy rod, which is then prepared by a rotated dipping process. To ensure a homogeneous chemical composition and prevent component segregation and high/low-density inclusions, the material was repeatedly melted 4 times in a vacuum arc melting furnace. During the rotated dipping process, the molten master alloy forms a convex molten pool under the action of surface tension, which provides the required molten metal for the wedge-shaped rim to dip, then the molten metal is rounded into metallic microfibers under the action of surface tension and its own gravity. However, a significant amount of residual stress is created inside the microfibers as a result of the fast condensation. Therefore, current annealing of the metallic microfibers is particularly important. Numerical calculations were performed using an amorphous heat conduction model with a heat source to obtain transient temperature rise curves. Comparing the glass transition temperature $T_g$ and the initial crystallization temperature $T_{x1}$ in the DSC curves and combining them with thermal physical parameters of the Co-based microfibers [30], it is found that as the temperature rises, the coercivity of the fibers increases. Thus, the AC current parameters were set to 60 mA - 240 mA with an annealing time or 8 min followed by air cooling. The specific process and test flow are depicted in Fig. 1.

A Setharam DSC 131 EVO differential scanning calorimeter was used to analyze the phase transition temperature of the metallic microfibers. The surface morphologies of the metallic microfibers were observed by scanning electron microscope (FEI QUANTA 650 FEG) and the elemental contents of the microfibers were determined by EDS. A JEOL JEM2010 transmission electron microscope was used to analyze the microscopic morphology, provide high-resolution images, and create selected area electron diffraction patterns of metallic microfibers. The hysteresis loops of the metallic microfibers were tested by a SQUID-VSM magnetic measurement system at room temperature with a timed magnetic field in the range of -5 T to 5 T, while the MI effect was tested by a comprehensive magneto-impedance test platform with a test frequency range of 5 Hz - 13 MHz.

The MI ratio $\Delta Z/Z_{max}$ and the magnetic field response sensitivity $\xi$ can be expressed as:

$$\frac{\Delta Z}{Z_{max}}(\%) = \left[\frac{Z(H_{ex}) - Z(H_{max})}{Z(H_{max})}\right] \times 100\% \tag{1}$$

$$\xi = \frac{d[\Delta Z/Z_{max}]}{dH_{ex}} \tag{2}$$

where $Z(H_{ex})$ stands for the impedance value ($\Omega$) under various external magnetic field conditions, $Z(H_{max})$ is the impedance value ($\Omega$) at the maximum external magnetic field strength, $H_{ex}$ represents the strength of the external magnetic field (Oe) provided by the Helmholtz coils, and $H_{max}$ is the maximum external magnetic field strength (Oe) provided by the Helmholtz coils.

## 3. Results and discussion

### *3.1 Analysis of the microstructure of CoFeSiBNb$_3$ metallic microfibers before and after AC current annealing*

The transient temperature characteristic curves of CoFeSiBNb$_3$ microfibers before and after AC current annealing are depicted in Fig. 2(a). As shown in the figure, the current amplitude became stable after increasing from 60 mA to 240 mA in the time range of $\Delta t = 0.1$ s to 0.2 s. Meanwhile, the temperatures inside the microfibers also show an upward trend. The ultimate temperatures of the microfibers at 60 mA, 100 mA, 140 mA, 190 mA, and 240 mA were 45.71 °C, 80.13 °C, 129.45 °C, 209.50 °C, and 311.87 °C, respectively. Figure 2(b) shows the DSC curves of the CoFeSiBNb$_3$ microfibers in the heating stage before and after AC current annealing. As can be seen, the microfibers did not crystallize at the temperature calculated by current values, indicating that the CoFeSiBNb$_3$ microfibers still have their typical amorphous structure after current annealing. The width of the subcooled liquid phase zone $\Delta T$ and the mixing enthalpy $\Delta H$ are relatively stable, indicating that the AC current annealed metallic microfibers have a good amorphous formation capacity and thermal stability, as shown in Table 1.

Figure 3 shows the SEM morphology and EDS spectra of CoFeSiBNb$_3$ microfibers before and after AC current annealing. CoFeSiBNb$_3$ microfibers are generally smooth and their surface is uniform and continuous with no noticeable macroscopic flaws. Their diameters range from 35 to 50 μm with satisfactory roundness. The elemental content changed relatively smoothly after AC annealing, as shown in Fig. 3(b) (d) (f) (h) (j) (l), which indicates that the appropriate intensity of current annealing did not alter the chemical composition of the microfibers. EDS spectroscopy revealed no significant shift in the composition and spectral position of the metallic microfibers after AC annealing.

The morphology from HRTEM of CoFeSiB amorphous microfibers is shown in Fig. 4(a). The figure shows that the metallic microfibers do not appear to be crystallized, and the electron diffraction pattern is characterized as an amorphous halo ring. It is evident from auto-correlation function (ACF) statistics that the microfibers are arranged with a degree of order of 0. Figure 4(b) shows the microscopic morphology of the metallic microfibers doped with 3 at.% Nb. As can be seen from the figure, the microstructure is homogeneous and the electron diffraction pattern is also characterized as an amorphous halo ring. The HRTEM pattern and autocorrelation results show that the metallic microfibers appear to possess less nano-ordered regions in the local micro-regions after Nb element doping, with an organizational orderliness $\psi$ of 4.6%. Figure 4(c) shows the HRTEM microscopic morphology following annealing with 140 mA. The degree of order in this chosen location is 29.69%, as seen from the image. The amorphous matrix as a whole seems to have more evident ordered micro-regions of atomic arrangement, where nano-clusters are present and more evenly distributed. When the Joule heat activation energy is modulated, residual stresses within the microfibers are removed, causing faster atom migration and diffusion in the ordered micro-regions and a more regular atomic arrangement. This results in a more regular atomic arrangement in the ordered micro-regions of the microfibers after AC current annealing compared to the as-prepared microfibers. The MI effect of the metallic microfibers may be considerably enhanced by the ordered micro-regions of atomic arrangement that develop after the modulation treatment as demonstrated by research [31].

### *3.2 Magnetic properties of CoFeSiBNb$_3$ microfibers before and after AC current annealing*

The hysteresis loops of CoFeSiBNb$_3$ microfibers are shown in Fig. 5 before and after AC current annealing. As observed in the figure, the magnetization intensity proliferates as the external magnetic field strength increases, then climbs more slowly until it reaches a particular value, and then rises even more slowly until it achieves saturation. When annealed at 190 mA, the metallic microfibers display exceptional general magnetic characteristics, with coercivity $H_c$, residual magnetization strength $M_r$, and permeability $\mu_m$ of 34.36 Oe, 13.94 emu/g, and 0.4326, respectively. The AC annealed metallic microfibers show a greater permeability, no discernible change in residual magnetization strength, lower coercivity, and a smaller area covered by the hysteresis loops than the as-prepared CoFeSiBNb$_3$ microfibers. After AC current annealing, the microfibers' overall magnetic properties are significantly improved due to a decrease in stress anisotropy and magneto-crystalline anisotropy brought on by forming atomically ordered micro-regions within the microfibers. This uniform distribution of magnetic anisotropy within the microfibers lowers the resistance to the domain wall displacement magnetization process. Thus, the microfibers' overall magnetic characteristics are greatly enhanced as the resistance to the displacement process is decreased.

Figure 6 shows the MI characteristic curves of the CoFeSiBNb$_3$ microfibers before and after AC current annealing. Figure 6(g) demonstrates that with increasing frequency, the maximum rate of change in impedance $[\Delta Z/Z_{max}]_{max}$ of AC annealed metallic microfibers tends to grow and then subsequently drop. Figure 6 (b) (d) (e) (f) illustrates the "single peak" MI curve for lower frequency current excitation with the magnetic field value at the peak (also known as the equivalent anisotropic field) $H_p = 0$ for the 60-mA AC annealed metallic microfibers at frequencies of 100 kHz, 1 MHz, and 3 MHz as well as 100 mA, 140 mA, 190 mA, and 240 mA annealed metallic microfibers at frequencies of 100 kHz and 1 MHz. This single peak phenomenon is primarily caused by domain wall displacement dominating the microfiber's magnetic domain structure at lower frequencies of excitation current and its high permeability. However, when the alternating current is applied along the axial direction of the metallic microfiber, cyclic magnetization is produced, and the cyclic permeability increases. The MI curve exhibits a "single peak" as the cyclic permeability falls with increasing magnetic field strength. Figure 6 (b) (d) (e) (f) also shows that the MI curve exhibits a "double peak" feature at 60 mA, 140 mA, 190 mA, and 240 mA annealed metallic microfibers at 5 MHz, 7 MHz, 10 MHz, and 13 MHz, respectively. This illustrates that the double peak characteristic exists over a broader frequency range and that the MI curve exhibits a double peak pattern at higher frequencies when $H_p$ is not zero. It is foremost that eddy current damping caused by the increase in frequency prevents the displacement of domain walls, and thus magnetic moment rotation dominates. During current annealing, $[\Delta Z/Z_{max}]_{max}$ rises from 169.69% at 60 mA to 216.81% at 140 mA before falling to 164.83% at 240 mA when $f$ =1 MHz. At the same time, the maximum value of the maximum impedance change $[\Delta Z/Z_{max}]_{max}$ for the fibers occurs at 1 MHz and 3 MHz as shown in Fig. 6 (b) (d) (f).

Figure 7 shows the curve of the magnetic field sensitivity of MI for CoFeSiBNb$_3$ microfibers before and after AC current annealing. As seen from Fig. 7(g), after homogenization, it is determined that the high-frequency characteristics of the metallic microfibers are more stable in the 100-mA current annealed state. It is evident that the sensitivity of the magnetic field response of the metallic microfibers varies at different frequencies after annealing at different currents. In contrast, with a maximum magnetic impedance change $[\Delta Z/Z_{max}]_{max}$, a maximum magnetic field sensitivity of MI $\xi_{max}$ of 216.81% and 31.04%/Oe, respectively, and the corresponding maximum magnetic field response field $H_k$ and the equivalent anisotropic field $H_p$ of 1 Oe and 2 Oe, respectively, the 140 mA current annealed state metallic microfibers have better MI characteristics at 1 MHz as shown in Fig. 7(d). It is thus demonstrated that AC current modulation treatment improves the MI characteristics of the metallic microfibers.

### *3.3. Mechanism analysis of magnetoimpedance characteristics improved by AC current annealing*

Figure 8 shows the mechanism behind the magnetic action of CoFeSiBNb$_3$ microfibers before and after AC current annealing. The figure shows that Joule heat is produced inside the metallic microfibers during current annealing, efficiently controlling the magnetic domain structure and eliminating residual strain. Meanwhile, the coalescing magnetic domains on the microfiber surface as a result of the doping of Nb elements in place of B elements causes an increase in the circumferential domain breadth and surface roughness. However, the formation of nanoclusters disproportionately impedes the domain wall displacement and makes the rotation of the magnetic moment more challenging. When the Nb element is over-doped, the ordered micro-areas of atoms are numerous and densely packed. As a result, the magnetostriction coefficient tends to be positive. Following current annealing, the current creates a stable circumferential magnetic field around the microfibers. This causes the magnetic domains to be distributed circumferentially on the microfiber surface with the external magnetic moments aligned parallel to the circumferential direction and the core uniformly distributed along the axial direction. Therefore, the microfibers' magnetic impedance effect is successfully controlled. During low annealing current, the stress relief allows for some improvement in the microfibers' circumferential domain structure, producing a small circumferential magnetic field and a small amount of Joule heat. But the effect is weak, so the magnetic impedance effect of the microfibers does not change significantly. When the current strength increases, the release of residual stress, interatomic migration, and diffusion are more fully completed. As ordered microregions of atomic arrangement appear, the toroidal magnetic field is simultaneously enhanced, the size of the toroidal magnetic domains of the microfibers is increased, and the toroidal permeability under low magnetic elasticity is increased. As a result, the magnetic anisotropy and sensitivity to magnetic field response are both increased. It should be noted that as the sizes of the ordered sections of the nanoclusters in the metallic microfibers grow too large, the magnetic domain structure inside the microfibers is disrupted, which reduces the impedance effect as the current strength rises. These changes also cause the material's soft magnetic characteristics to deteriorate.

## 4. Conclusions

In this study, the microstructure evolution during current annealing and its effect on the magnetic and MI properties was examined. The main conclusions are summarized as follows:

(1) Before and after the present annealing procedure, the surface of the CoFeSiBNb$_3$ microfibers was homogeneous and continuous with no noticeable geometrical alterations. The organizational orderliness of the metallic microfibers was enhanced after the annealing process, while their thermal stability and capacity for amorphous formation were somewhat diminished.

(2) Following AC current annealing, the general magnetic characteristics and MI of the CoFeSiBNb$_3$ microfibers exhibited a trend of rising and then dropping with increasing annealing current intensity. The 140-mA AC-annealed metallic microfibers, which has index values of 94.38 emu/g, 0.4326, 34.36 Oe, and 13.94 emu/g for $M_s$, $\mu_m$, $H_c$, and $M_r$, respectively, have reasonably outstanding magnetic characteristics. When $f$ = 1 MHz, the $[\Delta Z/Z_{max}]_{max}$, $\xi_{max}$, $H_k$, and $H_p$ values were 216.81%, 31.04%/Oe, 1 Oe, and 2 Oe, respectively.

(3) In the AC annealed states, more ordered microfiber structures were produced by the Joule heat released during the current annealing process, which also enhanced the atomic relaxation state forming micro-areas of atomic ordering. In addition to relieving stress, the current annealing procedure created a stable circumferential magnetic field that, when combined with the increased circumferential permeability and induced magnetic anisotropy of the microfibers, altered the structural distribution of the microfibers' magnetic domains and thus enhanced the MI effect.

**Acknowledgments**

This work was financially supported by the National Natural Science Foundation of China (NSFC) under grant nos. 52061035, 51871124, 51561026, and 51401111, Ministry of Education Fok Ying-tung Foundation for Young Teachers (no. 161043), "Grassland Talents"

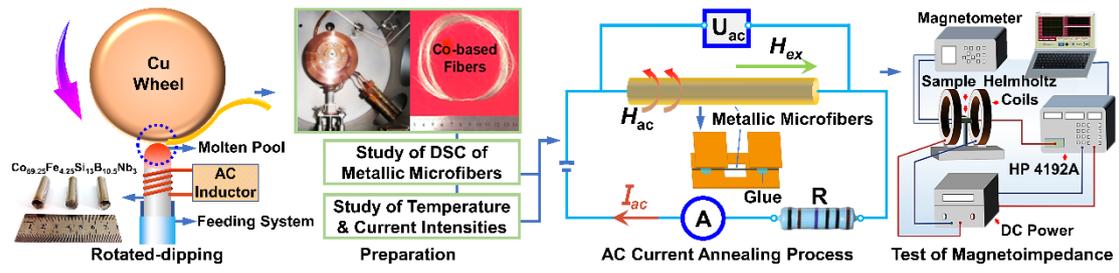

**Fig. 1.** CoFeSiBNb$_3$ microfibers' preparation and its testing flow chart schematic diagram.

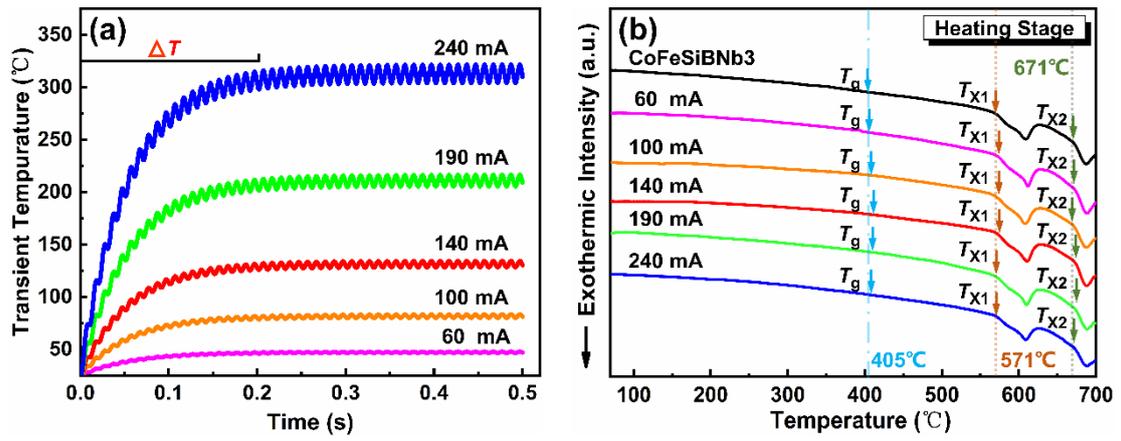

**Fig. 2**. Thermophysical parameters of CoFeSiBNb$_3$ microfibers before and after AC current annealing: (a) transient temperature characteristics curves and (b) DSC curve.

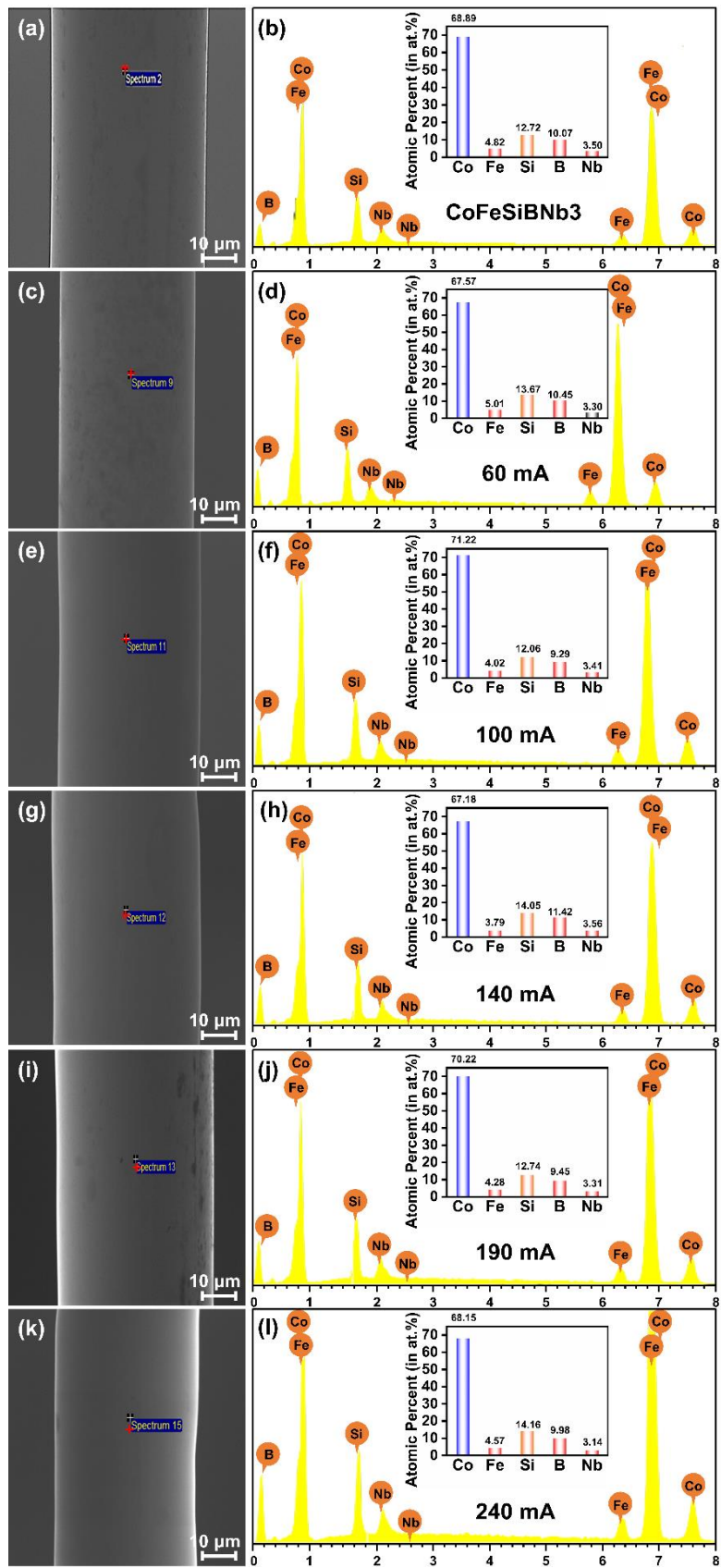

**Fig. 3.** SEM surface morphology and EDS analysis of CoFeSiBNb$_3$ microfibers before and after AC current annealing: (a), (b) CoFeSiBNb$_3$; (c), (d) 60 mA; (e), (f) 100 mA; (g), (h) 140 mA; (i), (j) 190 mA; and (k), (l) 240 mA.

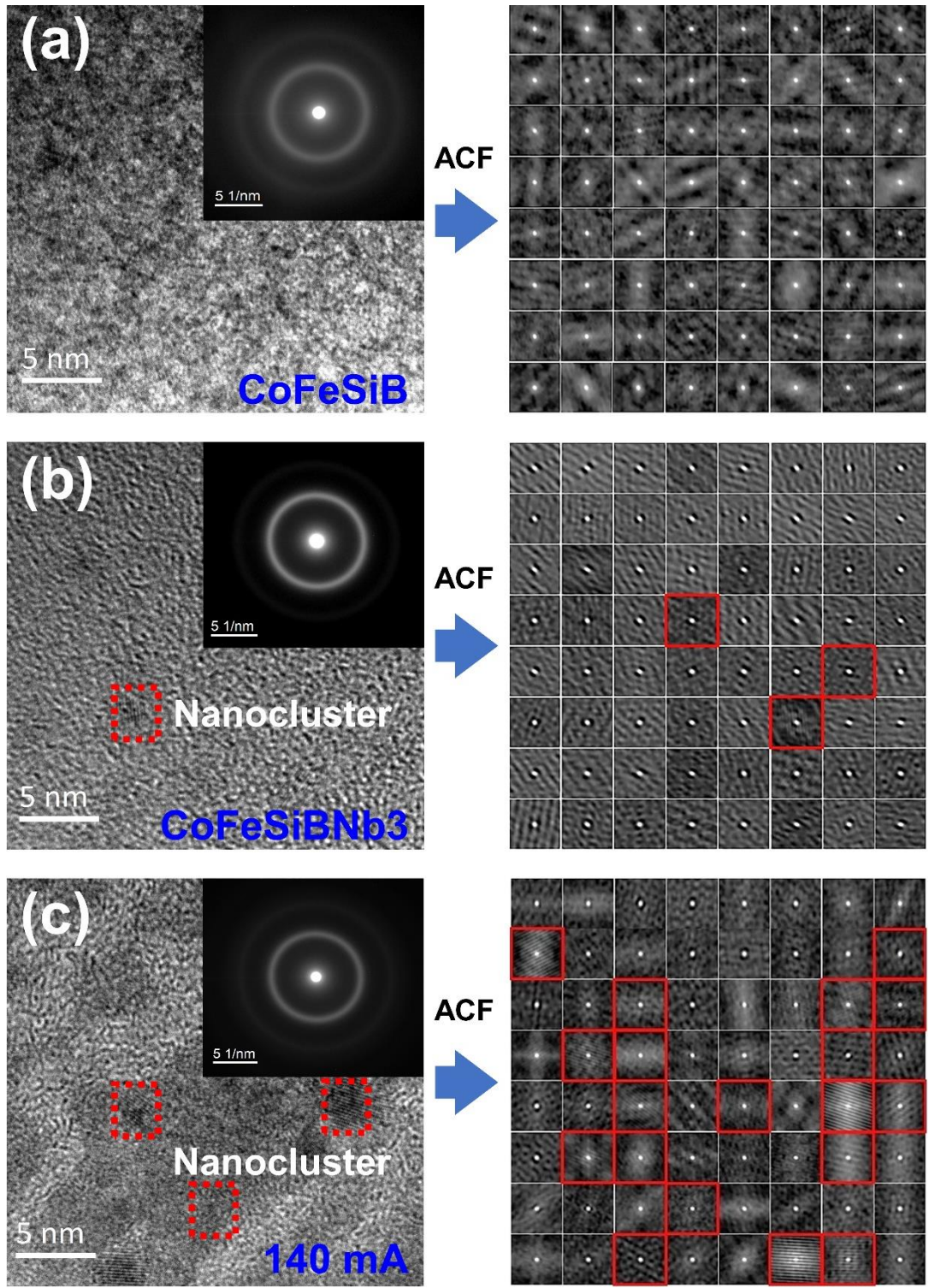

**Fig. 4.** HRTEM images, SAED, and ACF of metallic microfibers before and after AC current annealing: (a) CoFeSiB; (b) CoFeSiBNb$_3$; and (c) 140 mA.

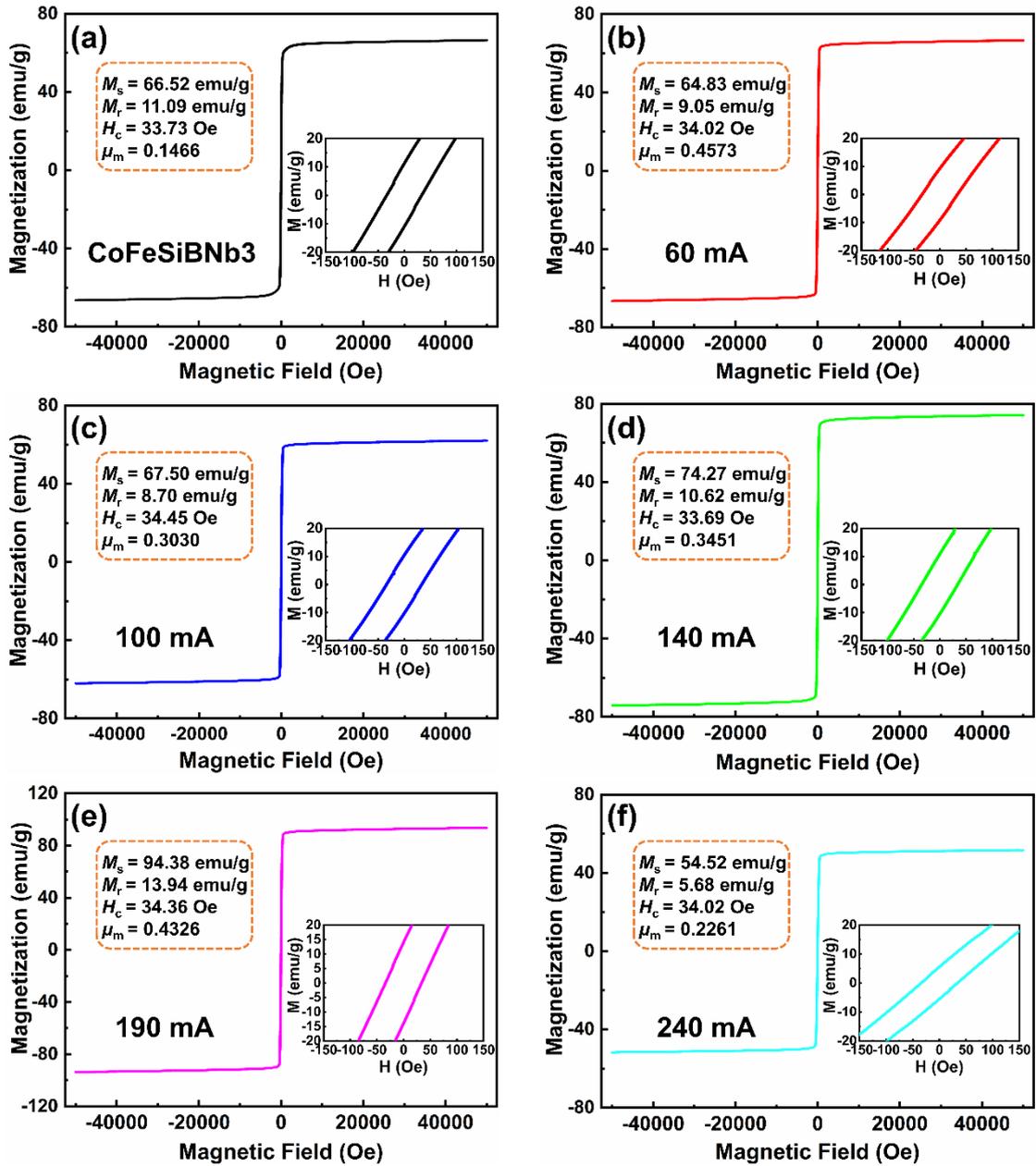

**Fig. 5.** Hysteresis loops of CoFeSiBNb$_3$ microfibers before and after AC current annealing: (a) CoFeSiBNb$_3$; (b) 60 mA; (c) 100 mA; (d) 140 mA; (e) 190 mA; and (f) 240 mA.

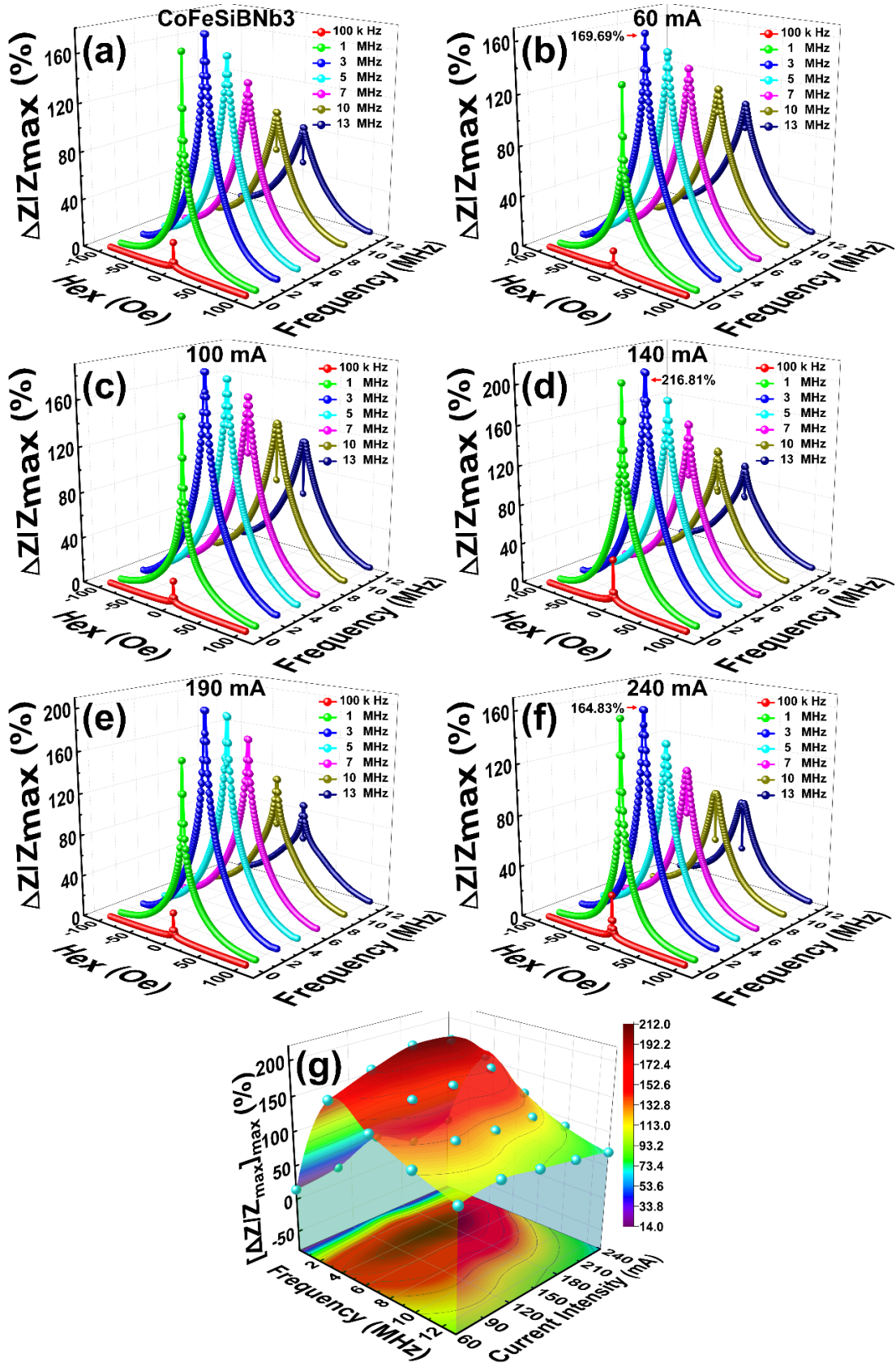

**Fig. 6.** MI characteristic curves of CoFeSiBNb$_3$ microfibers before and after AC current annealing: (a) CoFeSiBNb$_3$; (b) 60 mA; (c) 100 mA; (d) 140 mA; (e) 190 mA; (f) 240 mA; and (g) statistics of maximum MI ratio.

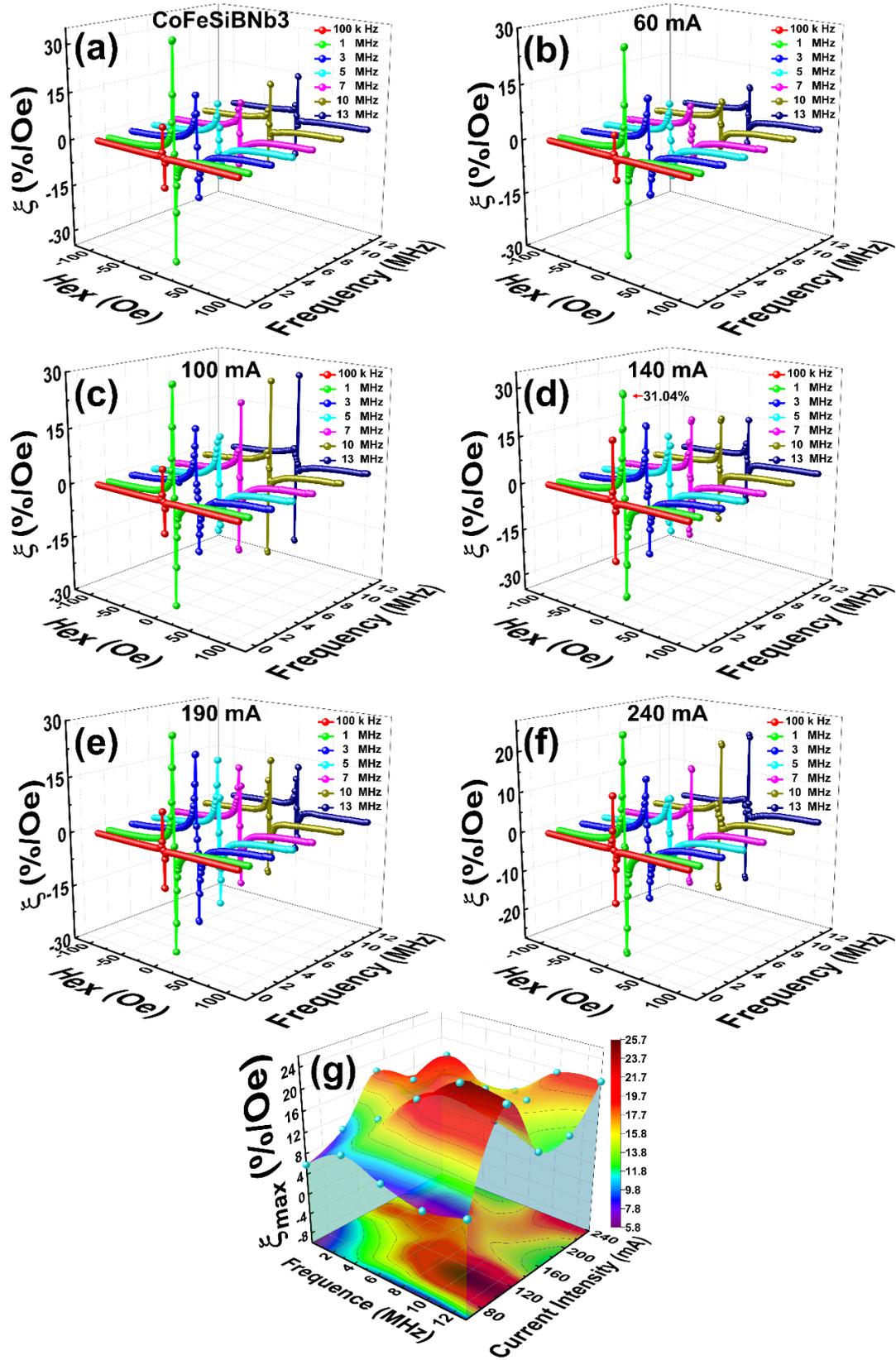

**Fig. 7.** Variation curves of maximum field sensitivity of MI $\xi_{max}$ of CoFeSiBNb$_3$ microfibers before and after AC current annealing: (a) CoFeSiBNb$_3$; (b) 60 mA; (c) 100 mA; (d) 140 mA; (e) 190 mA; (f) 240 mA; and (g) statistics of maximum field sensitivity of MI, $\xi_{max}$.

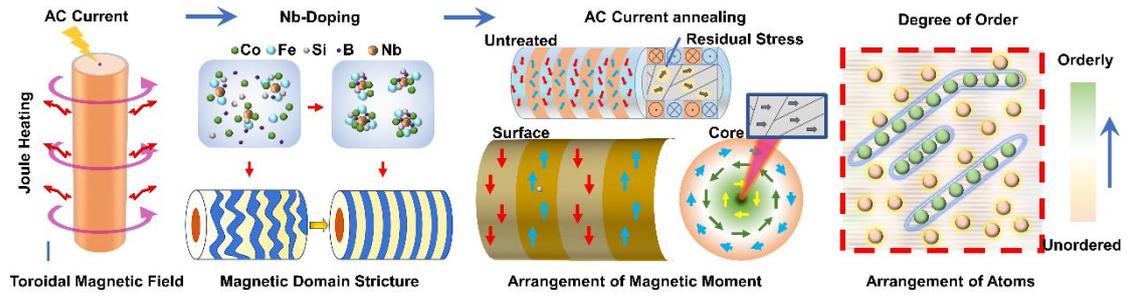

**Fig. 8.** Mechanisms of the magnetic orderings in CoFeSiBNb$_3$ metallic microfibers before and after annealing by AC current.

**Table 1.** Statistics of thermophysical parameters of CoFeSiBNb$_3$ microfibers before and after AC current annealing

| Current Annealing States | CoFeSiBNb$_3$ | 60 mA | 100 mA | 140 mA | 190 mA | 240 mA |
|---|---|---|---|---|---|---|
| $\Delta T$ (°C) | 166.5 | 168.3 | 167.0 | 165.8 | 162.9 | 163.7 |
| $\Delta H$ (J·g$^{-1}$) | -55.32 | -54.53 | -59.68 | -58.46 | -56.42 | -47.10 |